# Cooking the volumes

Amelia Carolina Sparavigna
Department of Applied Science and Technology
Politecnico di Torino, Torino, Italy

Cooking possesses a system of units of measurement, that includes measures of volumes based on pre-metric units. This paper discusses the cooking measures and compares their features with those of the ancient Roman measures of capacity.

**Introduction**
Besides being a necessity, cooking is a mixture of art and science. As a science, cooking possesses specific devices and instruments and units of measurement. And in fact, when we follow a recipe, accurate and consistent measures help in having the desired final result. As in any art, a little bit of uncertainties, or creativity, fulfils the personal taste and produces new effects. The recipes often allow this creativity, because do not specify all the quantities giving their measures. We find for instance the requirement of a "walnut-size lump" of butter, a "spoon" of sugar or a "sprinkle" of pepper. Generally, the relevant ingredients are proposed mainly with metric measures but pre-metric units are used too. Here I discuss the cooking measures, in particular of capacity, and compare their features with those of the ancient Roman measures.

**Cooking measures of capacity**
The recipes proposed in cookbooks have the ingredients measured in several ways. Usually, the solids, in particular the dry ingredients, are measured by weight in most of the world (for instance, 250 g flour, 100 g sugar). Volumes for dry ingredients are popular too, using the "cup" as a measure (two cups of flour, a cup of sugar). Sets of cups are available, having the size of 1/4, 1/8 of litre, and so on. The metric cup is in fact equal to 250 millilitres. For liquids, these measuring cups give accurate quantities. For solids, the use of cups is appropriate when an ingredient's exact measure is not necessary. When a spatula or ruler to level solids is used, the measuring cups sized to the top rim work well in obtaining the required ingredients. To increase the accuracy of the quantity of dry ingredients, a scale to weigh it is necessary.
Actually, the recipes use in most of the world the metric system of litres (millilitres), kilograms (grams) and degrees Celsius (°C). But the English-speaking world frequently measures weight in pounds, with volume measurement based on cooking utensils and pre-metric measures [1,2]. Some common volume measures are the Teaspoon, Dessertspoon, Tablespoon, Cup, Pint, Quart and Gallon (for the conversion to metric units, see [1,2]). Let us note that for the cooking system of measurement, the measures are classified as either for dry or for fluid ingredients. Some of these measures have similar names, but the actual measured volume is quite different.
Another feature of cooking measures is that some of them continue to be not decimal, but based on fraction system. For instance for USA measures we have that the Cup is 1/2 of a Pint, 1/4 of a Quart and 1/16 of Gallon [1]. Therefore, we see that pre-metric pre-decimal measures continue to be used nowadays, as they were before the introduction of the metric system. We find then in these cooking measures several features of the units of measurement that the Romans spread in their empire, measures that, due to their relevant social role, were ruled by laws.

**An Apicius' recipe**
Before discussing the Roman measures of capacity, let us note that the use of cookbooks and cooking systems of measures is quite old. In Europe, the earliest collection of recipes, compiled in the late 4th or early 5th century AD, is De re coquinaria, written in Latin by Apicius. The name of the author, probably a collective name, is associated with that of the Roman gourmet Marcus Gavius Apicius, lover of refined luxury who lived during the reign of Tiberius. Let me report one of the Apicius' recipes, that of the Tyropatinam.
"Accipies lac, quod adversus patinam aestimabis, tempera lac cum melle quasi ad lactantia, ova quinque ad sextarius mittis, si ad heminam, ova tria. In lacte dissulvis ita ut unum corpus facias, in Cumana colas et igni lento coques. Cum duxerit ad se, piper aspergis et inferes," that is, as proposed in its first English rendering [3], "Estimate the amount of milk necessary for this dish and sweeten it with honey to taste; to a pint (sextarius) of fluid take 5 eggs; for half a pint (ad heminam) dissolve 3 eggs in milk and beat well to incorporate thoroughly, strain through a colander into an earthen dish and cook on a slow fire [in hot water bath in oven]. When congealed sprinkle with pepper and serve". In this recipe, two volumes are given for milk: the Sextarius and the Hemina, which are measures of capacity of the ancient Rome.

**Roman measures of capacity**
The Roman system of measurement originated from the ancient measures used in the Mediterranean basin. Several books discuss the Roman measures. Let me suggest Ref.4, for instance. It is an interesting book having the specific references to the Latin literature where the measures are mentioned. The measure of capacity most frequently mentioned by Roman authors is the Amphora, called also Quadrantal, or Cadus. It has the volume of a cubic foot. An Amphora contains 2 Urnae, 8 Congii, 48 Sextarii, 96 Heminae, or Cotylae, as in Fig.1, which is reporting table VIII of Ref.5, for the system of fluid volumes. From this table we see clearly that the measures of capacity are based on a system of fractions. Being the Amphora a cubic Roman foot, its equivalence in metric units depends on the value of the Roman foot. Ref.4 is telling that the Amphora is nearly equal to 9 English Gallons, and the Sextarius to one English Pint and a half. For the system of the dry volumes, see Figure 2, reporting Table X of Ref.5. Here again we see that the set of measures is connected by fractions.
A Sextarius contained 2 Heminae, 4 Quartarii, 8 Acetabula and 12 Cyathi (see Fig.3, for a Sextarius [6] and an Acetabulum). A Cyathus was as much as one could easily swallow at once. It contained 4 Ligulae or Cochlearia spoonfuls.
The Congius, the eighth of an Amphora was equal to a cubic - half foot. This measure of oil or wine used anciently to be distributed by the magistrates or leading men among the people [4]. Hence, the Congiarium was a gratuity or largess of money, corn, or oil, given to the people, chiefly by the emperors, or a gratuity to the soldiers. The Congiaria of Augustus, from their smallness, used to be called Heminaria [4].
The Modius was the chief measure for dry materials, the third part of a cubic foot. Figure 4 shows the Modius of Ponte Puñide, Spain, which has an inscription telling that it is a MODII L(EX) IVXTA SACRAM IVSSI [...]EM DOMINORVM NOSTRORVM VALENTINIANI VALENTIS ET GRATIANI INVICTISSIMORVM PRINCIPVM etc. [7]. This container was then a standard, such as the Sextarius of Fig.4, which has the text on it telling ADDNNCC[AA] SEXTARIAL[US] EXACTUS [...] ITALICUS [...] etc. [6]. LEX IUXTA (according to the law) and EXACTUS (precise) indicate that these two objects have to be considered as historical standards, that is, standards defined by an authority, in particular the emperor.
Figures 5 and 6 show some Roman silver Cochlearia and Ligulae at the Museum of Palazzo Madama, Torino.

**Capacity and society**
As we can clearly understand, these measures of capacity had a relevant role among the people. The use of scales was not common: in the Roman markets goods, such as dry and liquid food, were sold by volumes. For instance, in the Leptis Magna's market, which is particularly well preserved, we can see the structure of some stalls, having marble counters with cylindrical holes corresponding to different measures of capacity. After sized to the top rim, the stall holder opened the bottom of the cylinder and the desired quantity fell in the basket of the customer [9,10]. As previously told, the Roman rulers stated the capacity of standards. Moreover, it is probable that each family had at home a Sextarius, such as that in Fig.3, to measure milk, oil, wine or beer, and a Modius for dry food.
Let me conclude my discussion reporting other information on the Roman habits as given in [4]. We see that the capacity measures were involved at the convivial tables too. It is told that the Romans set down the wine on the table, with the dessert. The wine was brought to the guests in earthen vases (amphorae, vel testae) with handles or in big-bellied jugs or bottles. The wine was mixed with water in a large vase or bowl, called crater, whence it was poured into cups, called calices or phialae. These cups were made of various materials, sometimes of silver and gold, beautifully engraved (see Fig.7). Usually handsome boys waited to mix the wine with water and serve it up; for which purpose they used a small goblet, a Cyathus, to measure it (see Fig.8, a delightful scene depicted in a Pompeii fresco, where Cupids are serving wine). The Cyathus, as given in Fig.2, was the twelfth part of a Sextarius. The cups used for convivial dinners were named according to the number of Cyathi, which they contained. We find then sextans, quadrans, triens, quincux and so on, cups. A sextans was a cup containing the sixth of the Sextarius, that is two Cyathi. A quadrans was a cup having a capacity equal to 1/4 of Sextarius, that is, of three Cyathi. A triens vel triental, contained 1/3 of Sextarius, equivalent to four Cyathi, quincunx for five Cyathi, etc.. As we can see from Fig.7, rich people had set of precious cups of different measures to display during banquets.
A last curiosity that we can find in Ref. 4 is the following. When the wine was brought on the table, the amphorae or the bottles had labels or small slips of parchments affixed on. These labels gave a short description of the quality and age of the wine. Nowadays, there is a sommelier or wine steward that, in fine restaurants, shows us the label and illustrates all the aspects of wine, sometimes using a tastevin. This small, very shallow silver cup is probably the evolution of the Roman Cyathus.

## ROMAN MEASURES OF CAPACITY.

### I. LIQUID MEASURES.

| | | | | | | | | |
|---|---|---|---|---|---|---|---|---|
| Ligula............................................................................... | | | | | | | | |
| 4 | Cyathus†......................................................................... | | | | | | | |
| 6 | 1½ | Acetabulum............................................................... | | | | | | |
| 12 | 3 | 2 | Quartarius, i. e., 1-4th of the *Sextarius*....................... | | | | | |
| 24 | 6 | 4 | 2 | Hemina or Cotyla................................................ | | | | |
| 48 | 12 | 8 | 4 | 2 | Sextarius, i. e., 1-6th of the *Congius*........... | | | |
| 288 | 72 | 48 | 24 | 12 | 6 | Congius......................................... | | |
| 1152 | 288 | 192 | 96 | 48 | 24 | 4 | Urna.................................... | |
| 2304 | 576 | 384 | 192 | 96 | 48 | 8 | 2 | Amphora Quadrantal.. |
| 46,080 | 11,520 | 7680 | 3840 | 1920 | 960 | 160 | 40 | 20 | Culeus........ |

Fig.1 This image reproduces the Table VIII of Ref.5, reporting the volume liquid measures.

## ROMAN MEASURES OF CAPACITY.

### II. DRY MEASURES.

| | | | | | | | |
|---|---|---|---|---|---|---|---|
| Ligula............................................................................... | | | | | | | |
| 4 | Cyathus†......................................................................... | | | | | | |
| 6 | 1½ | Acetabulum............................................................... | | | | | |
| 12 | 3 | 2 | Quartarius, i. e., 1-4th of the *Sextarius*....................... | | | | |
| 24 | 6 | 4 | 2 | Hemina or Cotyla................................................ | | | |
| 48 | 12 | 8 | 4 | 2 | Sextarius, i. e., 1-6th of the *Congius*........... | | |
| 384 | 96 | 64 | 32 | 16 | 8 | Semimodius................... | |
| 768 | 192 | 128 | 64 | 32 | 16 | 2 | Modius............... |

Fig.2 This image reproduces the Table X of Ref.5, reporting the volume dry measures.

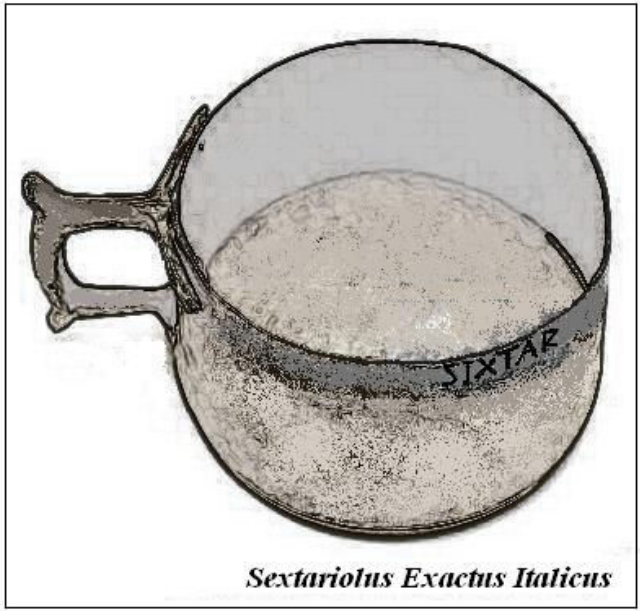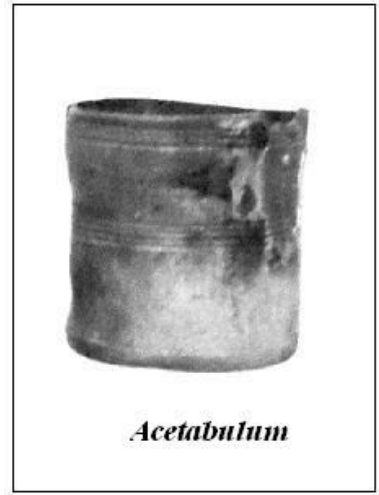

Fig.3 The Sextarius "exact measure" [6] and an Acetabulum (Egyptian Museum, Torino).

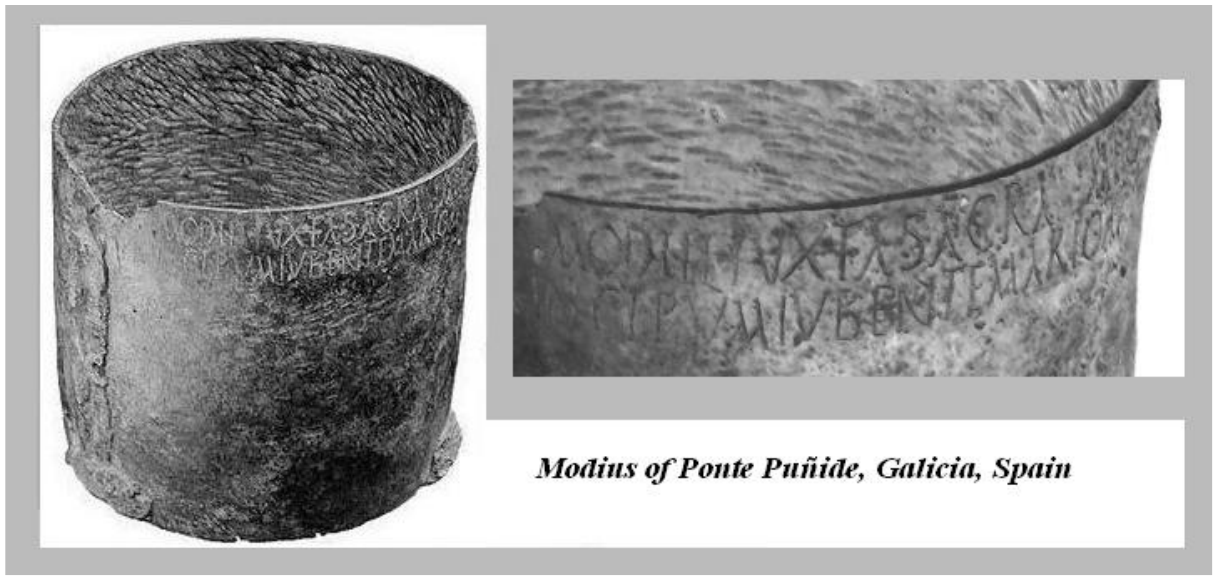

Fig.4 The Modius "exact measure" [8].

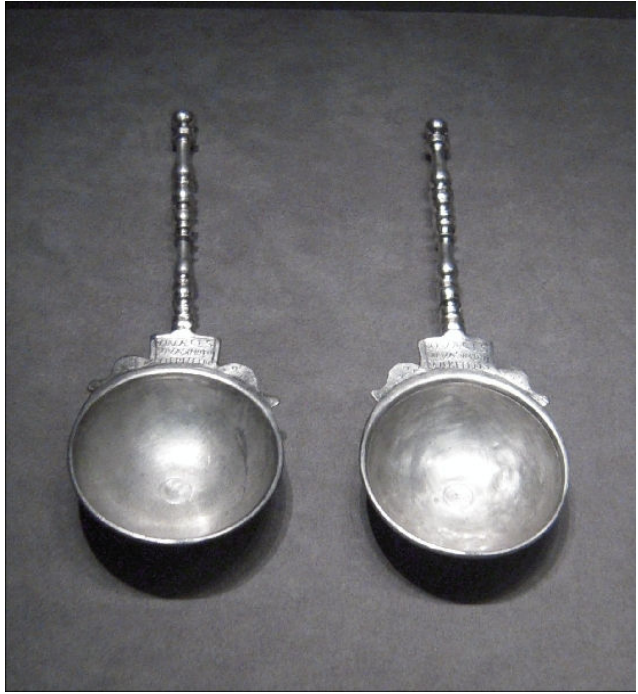

Fig.5 Roman Silver Cochlearia (Palazzo Madama, Torino)

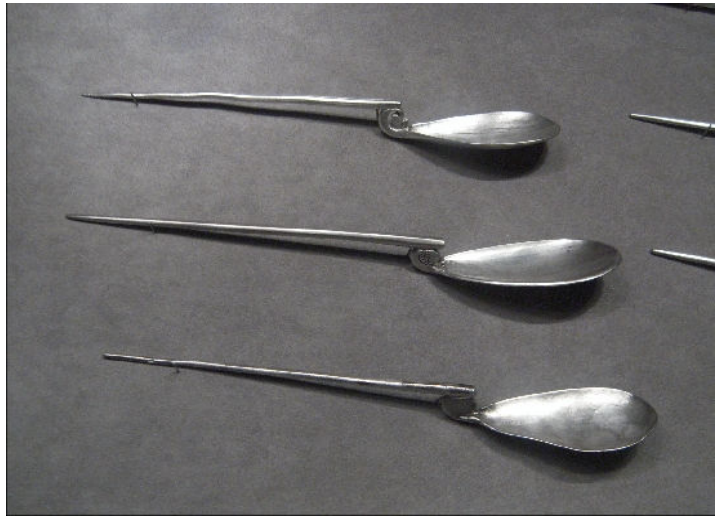

Fig.6 Roman Silver Ligulae (Palazzo Madama, Torino)

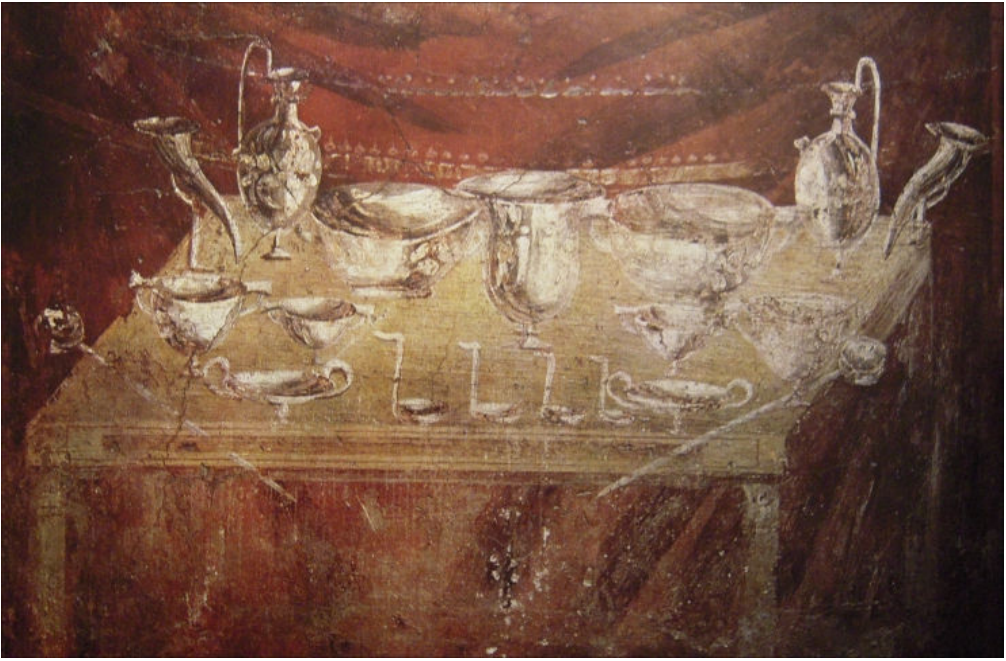

Fig.7 A beautiful table in a Pompeii fresco.

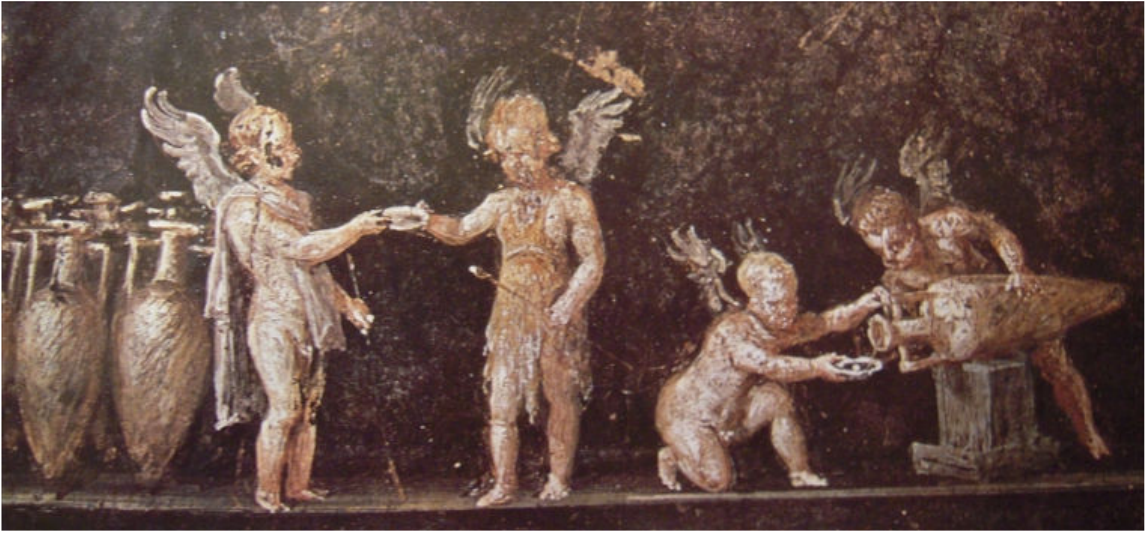

Fig.8 Cupids serving the wine from Amphorae in Cyathi, in a Pompeii fresco.